\documentclass[12pt,a4paper,final]{iopart}
\renewcommand{\figurename}{Figure}
\expandafter\let\csname equation*\endcsname\relax
\expandafter\let\csname endequation*\endcsname\relax

\usepackage{amsfonts}
\usepackage{verbatim} 
\usepackage[switch]{lineno}
\usepackage{amsmath, amssymb, amsthm}
\usepackage{epsfig, color}
\usepackage{systeme}
\usepackage{multirow, comment, enumerate}
\usepackage{enumitem}
\usepackage{esint}
\usepackage{lipsum}
\usepackage{bm}
\usepackage{bbm}
\usepackage{hyperref}

\allowdisplaybreaks

\definecolor{myblue}{RGB}{0,50,200}
\hypersetup{
colorlinks,
citecolor=myblue,
linkcolor=myblue,
urlcolor=myblue
}
\definecolor{mycyan}{RGB}{45, 99, 135}
\definecolor{mypink}{cmyk}{0, 0.7808, 0.4429, 0.1412}

\newcommand{\bms}{\mathfrak}
\newcommand{\mca}{\mathcal}
\newcommand{\mbb}{\mathbb}
\newcommand{\mrm}{\mathrm}
\newcommand{\avg}[1]{\langle #1\rangle}

\newcommand{\avgl}[1]{\left\langle #1\right\rangle}
\newcommand{\set}[1]{\lbrace #1\rbrace}
\newcommand{\bra}[1]{\left( #1 \right)}
\newcommand{\bras}[1]{\left[ #1 \right]}
\newcommand{\brab}[1]{\left\{ #1 \right\}}

\newcommand{\bX}{\mca{X}}
\newcommand{\bM}{\mca{M}}
\newcommand{\bE}{\mca{E}}
\newcommand{\bZ}{\mca{Z}}
\newcommand{\bO}{\mca{O}}
\newcommand{\bXr}{\mca{X}^\dagger}
\newcommand{\bMr}{\mca{M}^\dagger}
\newcommand{\bZr}{\mca{Z}^\dagger}
\newcommand{\Pe}{\mrm{Pe}}
\newcommand{\pp}{\partial}
\newcommand{\intf}{\int_{-\infty}^{\infty}}
\newcommand{\inth}{\int_{0}^{\infty}}
\newcommand{\csch}{\mrm{csch}}

\definecolor{mygreen}{RGB}{20, 150, 70}

\begin{document}

\title[Uncertainty relation under information measurement and feedback control]{Uncertainty relation under information measurement and feedback control}

\author{Tan Van Vu and Yoshihiko Hasegawa}

\address{Department of Information and Communication Engineering, Graduate School of Information Science and Technology, The University of Tokyo, Tokyo 113-8656, Japan}

\eads{\mailto{tan@biom.t.u-tokyo.ac.jp} and \mailto{hasegawa@biom.t.u-tokyo.ac.jp}}

\date{\today}

\begin{abstract}
Here, we investigate the uncertainty of dynamical observables in classical systems manipulated by repeated measurements and feedback control; the precision should be enhanced in the presence of an external controller but limited by the amount of information obtained from the measurements.
We prove that the entropy production and the information quantity constrain from below the fluctuation of arbitrary observables that are antisymmetric under time reversal.
The information term is the sum of the mutual entropy production and the Kullback--Leibler divergence, which characterises the irreversibility of the measurement outcomes.
The result holds for finite observation times and for both continuous- and discrete-time systems.
We apply the derived relation to study the precision of a flashing Brownian ratchet.
\end{abstract}

% \pacs{05.70.Ln}{Nonequilibrium and irreversible thermodynamics}
% \pacs{05.40.-a}{Fluctuation phenomena, random processes, noise, and Brownian motion}
% \pacs{05.20.-y}{Classical statistical mechanics}
% PACS, the 

\section{Introduction}

Substantial progresses have been made in stochastic thermodynamics (ST) over the past two decades, resulting in a comprehensive theoretical framework for the study of small systems \cite{Seifert.2012.RPP,Broeck.2015.PA}.
ST investigates the physical properties of nonequilibrium systems and has a wide range of applications in both physics and biology \cite{Seifert.2012.RPP}.
One of its main results is the fluctuation theorem \cite{Gallavotti.1995.PRL,Jarzynski.1997.PRL}, which expresses universal properties relevant to the symmetry of the probability distributions of thermodynamic quantities such as heat, work and entropy production.

A trade-off between the precision of the currents and the thermodynamic cost, known as the thermodynamic uncertainty relation (TUR), has been recently reported \cite{Barato.2015.PRL,Gingrich.2016.PRL,Pietzonka.2016.PRE,Polettini.2016.PRE,Horowitz.2017.PRE,Dechant.2018.JSM}.
In general, the TUR states that, at a finite time in steady-state systems, the relative fluctuation of the arbitrary currents is lower bounded by the reciprocal of the total entropy production; in other words, it asserts the impossibility of attaining high precision without increasing the thermal cost.
Its original form is
\begin{equation}\label{eq:TUR.conv}
\frac{\mrm{Var}[\mca{O}]}{\avg{\mca{O}}^2}\ge\frac{2}{\avg{\sigma}},
\end{equation}
where $\mca{O}$ is a time-integrated current, $\avg{\mca{O}}$ and $\mrm{Var}[\mca{O}]$ are the current mean and variance and $\avg{\sigma}$ is the average entropy production.

Numerous studies have been conducted on TUR, including extensions to discrete-time Markov jump processes \cite{Proesmans.2017.EPL,Chiuchiu.2018.PRE}, periodically driven dynamics \cite{Barato.2018.NJP,Koyuk.2019.JPA}, multidimensional observables \cite{Dechant.2018.JPA}, underdamped \cite{Vu.2019.PRE.Underdamp} and time-delayed Langevin dynamics \cite{Vu.2019.PRE.Delay} and information-theoretic \cite{Hasegawa.2019.PRE} and hysteretic bounds \cite{Proesmans.2019.JSM}; they involve applications in biochemical systems \cite{Barato.2015.JPCB,Patrick.2016.JSM,Hwang.2018.JPCL}, heat engine efficiency \cite{Pietzonka.2018.PRL,Holubec.2018.PRL} and a range of specific problems \cite{Falasco.2016.PRE,Rotskoff.2017.PRE,Garrahan.2017.PRE,Gingrich.2017.PRL,Hyeon.2017.PRE,Brandner.2018.PRL,Manikandan.2018.JPA,Terlizzi.2018.JPA,Carollo.2019.PRL,Shreshtha.2019.EPL}.
In a recent work \cite{Hasegawa.2019.PRL}, we have demonstrated that a generalised TUR can be derived from the fluctuation theorem, showing an intimate connection among these universal relations.

Feedback control by an external protocol that depends on the measurement outcome is ubiquitous in both physics and biology and plays important roles in the study of nonequilibrium systems.
The thermodynamics of feedback control \cite{Touchette.2000.PRL,Touchette.2004.PA,Sagawa.2008.PRL,Sagawa.2010.PRE,Horowitz.2011.EPL,Sagawa.2012.PRE,Hartich.2014.JSM,Barato.2014.PRL,Parrondo.2015.NP,Potts.2018.PRL} provides a crucial framework for analysing systems in the presence of Maxwell's demon, which can extract work from the system beyond the limit set by the conventional second law.
The system performance can be significantly enhanced by applying the measured information about itself; moreover, such information could improve the precision of observables such as the displacement of a molecular motor \cite{Serreli.2007.N,Lau.2017.MH}.
This leads us to ask how the relative fluctuation of arbitrary observables is constrained in the presence of feedback control.

Here, we study TUR for steady-state systems involving repeated measurements and feedback control; in particular, we define a lower bound on the fluctuation of arbitrary dynamical observables that are antisymmetric under time reversal.
We prove that $\mrm{Var}[\bO]/\avg{\bO}^2$, where $\avg{\bO}$ and $\mrm{Var}[\bO]$ are respectively the mean and the variance of the arbitrary observable $\bO$, is lower bounded by a function of $\avg{\sigma}$ [cf.~Eq.~\eqref{eq:sigma.definition}], which denotes a quantity reflecting the thermal cost and mutual entropy production.
Due to the information flow, the observable fluctuation is bounded from below not only by the thermal energy consumed in the system but also by the quantity of information obtained from the external controller.
The inequality is valid for arbitrary observation times and for discrete- or continuous-time systems because the derivation does not require the underlying dynamics.
In addition, for Langevin dynamics involving continuous measurement and feedback control, we provide a tighter bound on the fluctuation of time-integrated currents.

Then, we apply the results to the study of a flashing ratchet \cite{Schimansky.1997.PRL,Freund.1999.PRE,Cao.2004.PRL,Craig.2008.AP}, in which the asymmetric potential is switched between on and off to induce a directed motion.
The investigated device is a nonequilibrium Brownian ratchet, which has been applied for modelling biological processes such as actin polymerisation \cite{Mogilner.1999.EBJ} and ion transportation \cite{Siwy.2002.PRL}.
To analyse the observable uncertainty under feedback control, we consider a flashing ratchet using imperfect information about the system state to rectify the motion of a diffusive particle; the ratchet acts as a Maxwell's demon by utilising the measured information to maximise the instant velocity.
Besides the mean velocity, which is the most common quantity used for transport characterisation, the relative fluctuation of the displacement, reflecting its precision, is another important attribute.
We also empirically verify the derived bound for the displacement of both discrete- and continuous-state ratchets.
To the best of our knowledge, this is the first time that a lower bound on the precision of an information ratchet is provided.

\section{Model}

We consider a classical Markovian system manipulated by repeated feedback, where an external controller uses the acquired information to evolve the system.
The system state is measured along a trajectory and the outcome is utilised to update the control protocol.
The time and state space of the system can be discrete or continuous.
We assume that every transition is reversible, i.e. if the transition probability from state $x$ to state $x'$ is nonzero, the reversed transition probability from $x'$ to $x$ is positive.

Suppose that we observe the system during a time interval $[0,\mca{T}]$.
Let $\bX$ and $\bM$ be the trajectories of its states and measurement outcomes, respectively; their time-reversed counterparts are therefore denoted as $\bXr$ and $\bMr$.
Then, we define the following trajectory-dependent quantity:
\begin{equation}
\sigma(\bX,\bM)\equiv\ln\frac{\mca{P}_{\rm F}(\bX,\bM)}{\mca{P}_{\rm R}(\bXr,\bMr)},\label{eq:sigma.definition}
\end{equation}
where $\mca{P}_{\rm F}(\bX,\bM)$ and $\mca{P}_{\rm R}(\bXr,\bMr)$ are the joint probabilities of observing trajectories in the forward and time-reversed processes, respectively.
As shown latter, the quantity $\avg{\sigma}$ constrains the fluctuation of the time-antisymmetric observables.
We can easily confirm that $\sigma$ satisfies the integral fluctuation theorem
\begin{equation}
\avg{e^{-\sigma}}=1.\label{eq:sigma.IFT}
\end{equation}
By applying the Jensen inequality to Eq.~\eqref{eq:sigma.IFT}, we can readily obtain $\avg{\sigma}\ge 0$.
This inequality can be considered as the second law of thermodynamics for a full system (e.g. system and controller), while $\sigma$ can be identified as its total entropy production.

\subsection{Discrete measurement and feedback control}

Now, let us apply the measurement and feedback control to the system from time $t=0$ up to the time $t=\mca{T}$.
For simplicity, we define $t_0\equiv 0,~t_{N}\equiv\mca{T}$.
Suppose that we perform measurements discretely at the predetermined times $t_0,t_1,\dots,t_{N-1}$ and the measurement outcomes are $\bM=\{m_0,m_1,\dots,m_{N-1}\}$.
The measurement times are $t_i=i\Delta t~(i=0,\dots,N)$, where $\Delta t=\mca{T}/N$ denotes the time gap between two consecutive measurements.
Let $\bX=\{x_0,x_1,\dots,x_N\}$ be the system states during the control process, where $x_i$ denotes the system state at time $t=t_i$ for each $i=0,\dots,N$; then, the measurement and feedback schemes are as follows.
First, at the time $t=t_0$, the observable is measured with outcome $m_0$, and the system is then driven with the protocol $\lambda(m_0)$ from $t=t_0$ to $t=t_1$.
At each subsequent time $t=t_{i}~(i=1,\dots,N-1)$, the measurement is performed and the corresponding outcome is $m_{i}$.
The protocol is immediately changed from $\lambda(m_{i-1})$ to $\lambda(m_i)$, and it remains constant until the time $t_{i+1}$.
The procedure is repeated up to the time $t=t_{N}$, which ends with the protocol $\lambda(m_{N-1})$.
Herein, it is assumed that the time delay required for measuring and updating the protocol can be ignored.
The measurement error is characterised using a conditional probability $p(m_k|x_k)$, where $x_k$ denotes the actual system state, while $m_k$ denotes the measurement outcome at time $t_k$.
This implies that the outcome depends on only the system's state immediately before the measurement.
In the sequel, we assume that the system is in the steady state under the measurement and feedback control.

Following Ref.~\cite{Kundu.2012.PRE}, we consider a time-reversed process in which $\bXr=\{x_N,\dots,x_0\},\bMr=\{m_{N-1},\dots,m_0\}$ and the measurements are performed at times $t_i^{\dagger}=\mca{T}-t_{N-i}$ for each $i=0,\dots,N-1$.
Since the control protocol is time-independent (i.e. it depends only on the measurement outcome), we have $\mca{P}_{\rm F}=\mca{P}_{\rm R}\equiv\mca{P}$.
Taking the ratio of the probabilities of the forward path and its conjugate counterpart, we obtain \cite{Kundu.2012.PRE}
\begin{equation}
\sigma=\ln\frac{\mca{P}(\bX,\bM)}{\mca{P}(\bXr,\bMr)}=\Delta s+\Delta s_{\rm m}+\Delta s_{\rm i},\label{eq:DFT.path}
\end{equation}
where each term in the right-hand side of Eq.~\eqref{eq:DFT.path} is expressed as follows:
\begin{equation}\label{eq:ent.def3}
\begin{aligned}
\Delta s&=\ln\frac{P^{\rm ss}(x_0)}{P^{\rm ss}(x_N)},\\
\Delta s_{\rm m}&=\ln\bras{\prod_{i=0}^{N-1}\frac{w(x_{i+1},t_{i+1}|x_i,t_i,m_i)}{w(x_{i},\mca{T}-t_i|x_{i+1},\mca{T}-t_{i+1},m_{i})}},\\
\Delta s_{\rm i}&=\ln\bras{\prod_{i=0}^{N-1}\frac{p(m_i|x_i)}{p(m_{i}|x_{i+1})}},
\end{aligned}
\end{equation}
where $P^{\rm ss}(x)$ denotes the steady-state distribution of the system and $w(x',t'|x,t,m)$ denotes the transition probability.
The first term $\Delta s$ and second term $\Delta s_{\rm m}$ denote the change in the system entropy and the medium entropy, respectively.
The last term $\Delta s_{\rm i}$ involves the probability that characterises the error in measurements; thus, it can be considered as an information quantity.
When $p(m|x)$ is the same uniform distribution for all $x$, the measurement is completely random and does not provide any valuable information.
In this case, $\Delta s_{\rm i}=0$, which shows that the system does not obtain any information from measurements.
Since $\avg{\Delta s+\Delta s_{\rm m}+\Delta s_{\rm i}}\ge0$, we have $\avg{\Delta s+\Delta s_{\rm m}}\ge-\avg{\Delta s_{\rm i}}$.
This implies that the entropy production of the system can be negative because of the effect of measurement and feedback control.

Next, we derive a lower bound on the fluctuation of the arbitrary dynamical observables that are antisymmetric under time reversal.
In particular, we focus on a bound for $\mrm{Var}[\bO]/\avg{\bO}^2$, where $\bO$ satisfies the antisymmetric condition $\bO[\bXr]=-\bO[\bX]$.
Current-type observables always satisfy this condition. 

By considering $P(\sigma)$ as the probability distribution of $\sigma$, i.e. $P(\sigma)=\int\mca{D}\bX\mca{D}\bM\,\delta(\sigma-\sigma(\bX,\bM))\mca{P}(\bX,\bM)$, we can show that $\sigma$ satisfies the strong detailed fluctuation theorem
\begin{equation}
\frac{P(\sigma)}{P(-\sigma)}=e^\sigma.
\end{equation}
We have previously demonstrated that a generalised TUR can be derived from the detailed fluctuation theorem (DFT) \cite{Hasegawa.2019.PRL}.
This derivation does not require detailed underlying dynamics and can be flexibly applied to other systems if the strong DFT is valid.
Based on Ref.~\cite{Hasegawa.2019.PRL}, we can prove that the observable fluctuation is bounded from below by a term involving $\avg{\sigma}$ as follows:
\begin{equation}
\frac{\mrm{Var}[\bO]}{\avg{\bO}^2}\ge\csch^2\bras{f\bra{\frac{\avg{\sigma}}{2}}}=\csch^2\bras{f\bra{\frac{\avg{\Delta s+\Delta s_{\rm m}}+\avg{\Delta s_{\rm i}}}{2}}},\label{eq:main.UR}
\end{equation}
where $f(x)$ denotes the inverse function of $x\tanh(x)$.
This lower bound is analogous to that utilised in Ref.~\cite{Timpanaro.2019.PRL}, where the TUR was derived from exchange fluctuation theorems for heat and particle exchange between multiple systems.
In addition to the system entropy production $\avg{\Delta s+\Delta s_{\rm m}}$, the information term, $\avg{\Delta s_{\rm i}}$, also appears in the bound.
Inequality \eqref{eq:main.UR} demonstrates that the precision of the arbitrary observables is constrained not only by the entropy production but also by the information obtained from the system measurement.
Since $\csch^2(f(x))$ is a decreasing function of $x$, the fluctuation of observables is reduced when the information quantity increases.
If we consider the limit $\Delta t\to 0$, i.e. the time gap between two consecutive measurements vanishes, the measurement and feedback control become continuous; therefore, this bound is also valid for the continuous-measurement case if $\avg{\sigma}$ is well defined in such limit (by properly constructing trajectories $\bX$ and $\bM$ and their corresponding time-reversed counterparts $\bXr$ and $\bMr$).
The detailed derivation of Eq.~\eqref{eq:main.UR} is provided in \ref{app:deriv.TUR.disc}.

Since $\csch^2[f(\avg{\sigma}/2)]\ge2/(e^{\avg{\sigma}}-1)$, the bound in Eq.~\eqref{eq:main.UR} is tighter than that in Ref.~\cite{Proesmans.2017.EPL}, where the TUR is derived for discrete-time Markovian processes in the long-time limit.
However, the derived bound is not as tight as the conventional bound $2/\avg{\sigma}$.
This happens because the derived inequality holds for both continuous- and discrete-time systems, while the conventional bound is valid only for the formers \cite{Shiraishi.2017.arxiv}.
In Section \ref{sect:exam}, we show that the conventional bound is actually violated for a discrete-time model.

\subsection{Continuous measurement and feedback control}

Here, we discuss systems involving continuous measurement and feedback control; in particular, we consider a Langevin system whose state variable is $x$.
For simplicity, the system is assumed to be one-dimensional.
The extension to multidimensional systems is straightforward.
The system dynamics is governed by the following equation:
\begin{equation}
\dot{x}=f(x,m)+\xi,\label{eq:Langevin.feedback}
\end{equation}
where the dot indicates the time derivative and $f(x,m)$ is the force depending on $x$ and the measurement outcome $m$.
$\xi$ is the zero-mean Gaussian white noise with variance $\avg{\xi(t)\xi(t')}=2D_{x}\delta(t-t')$, where $D_{x}$ denotes the noise intensity of $\xi$.

The measurement error is commonly incorporated by adding another zero-mean Gaussian noise, $e$, to the read-out of $x$, i.e. $m=x+e$.
Since the white noise fluctuations are violent, we assume that $e$ is a coloured noise.
Specifically, $e$ is modelled by the Ornstein--Uhlenbeck (OU) process as
\begin{equation}
\dot{e}=-e+\eta,\label{eq:OU.proc}
\end{equation}
where $\eta$ is the zero-mean Gaussian white noise with variance $\avg{\eta(t)\eta(t')}=2D_e\delta(t-t')$.
$D_e$ denotes the noise intensity of $\eta$ and also reflects the magnitude of the measurement error.
To ensure a clear illustration, we use the OU process to model the measurement error; however, another modelling of the noise does not affect the result as long as the noise is modelled by equilibrium overdamped Langevin dynamics (i.e. when the dynamics of $e$ is described as $\dot{e}=g(e)+\eta$, where $g(e)$ denotes a proper function).
In the sequel, we show that the analytical form of $\avg{\sigma}$ is independent of the form of $g(e)$.

Now, to investigate the analytical form of $\avg{\sigma}$, let us discretise the problem and take the continuous-time limit at the end.
Let $\bX=[x_0,x_1,\dots,x_N]$ and $\bM=[m_0,m_1,\dots,m_{N-1}]$ be the trajectories of the system states and measurement outcomes, respectively, in the forward process.
Here, $x_i\equiv x(i\Delta t)$, $m_i\equiv m(i\Delta t)$ and $\Delta t=\mca{T}/N$.
Their time-reversed counterparts in the backward process are $\bXr=[x_N,x_{N-1},\dots,x_0]$ and $\bMr=[m_N,m_{N-1},\dots,m_1]$.
Defining $e_i\equiv m_i-x_i$, the joint path probabilities, $\mca{P}(\bX,\bM)$ and $\mca{P}(\bXr,\bMr)$, can be expressed as follows:
\begin{align}
\mca{P}(\bX,\bM)\propto P^{\rm ss}(x_0,m_0)&\exp\bra{-\sum_{i=0}^{N-1}\frac{\bras{x_{i+1}-x_i-f(x_i,m_i)\Delta t}^2}{4D_{x}\Delta t}}\nonumber\\
&\times\exp\bra{-\sum_{i=1}^{N-1}\frac{\bras{e_i-e_{i-1}(1-\Delta t)}^2}{4D_e\Delta t}},\label{eq:forw.path.prob} \\
\mca{P}(\bXr,\bMr)\propto P^{\rm ss}(x_N,m_N)&\exp\bra{-\sum_{i=0}^{N-1}\frac{\bras{x_i-x_{i+1}-f(x_{i+1},m_{i+1})\Delta t}^2}{4D_{x}\Delta t}}\nonumber \\
&\times\exp\bra{-\sum_{i=2}^{N}\frac{\bras{e_{i-1}-e_i(1-\Delta t)}^2}{4D_e\Delta t}}.\label{eq:back.path.prob}
\end{align}
Using Eqs.~\eqref{eq:forw.path.prob} and \eqref{eq:back.path.prob} and taking the $\Delta t\to 0$ limit, we then obtain
\begin{equation}
\avg{\sigma}=\avgl{\ln\frac{\mca{P}(\bX,\bM)}{\mca{P}(\bXr,\bMr)}}=\frac{1}{D_{x}}\avgl{\int_0^{\mca{T}}dt\,f(x,m)\circ\dot{x}},\label{eq:sigma.Langevin}
\end{equation}
where $\circ$ denotes the Stratonovich product.
This expression can be used for the numerical evaluation of $\avg{\sigma}$.
We note that $\avg{\sigma}$ is a limit of the discrete sum $1/D_x\sum_{i}(x_{i+1}-x_i)[f(x_{i+1},m_{i+1})+f(x_i,m_i)]/2$.
When $f(x,m)$ is differentiable, $\avg{\sigma}$ is equivalent to the limit of the sum $1/D_x\sum_{i}(x_{i+1}-x_i)f((x_{i+1}+x_i)/2,(m_{i+1}+m_i)/2)$.
However, when $f(x,m)$ is non-differentiable, it is not the case and the former should be employed.

Let us investigate the information contribution of $\avg{\sigma}=\mca{T}\bar{\sigma}$, where $\bar{\sigma}$ denotes the total entropy production rate.
We can decompose $\bar{\sigma}$ into two positive terms as follows \cite{Horowitz.2015.JSM,Horowitz.2014.PRX}:
\begin{equation}
\bar{\sigma}=\bar{\sigma}_x+\bar{\sigma}_e,
\end{equation}
where $\bar{\sigma}_z=\bar{S}_z+\bar{Q}_z/D_z\ge 0$ denotes the entropy production rate contributed from $z$, for each $z\in\{x,e\}$.
Here, $\bar{S}_z=-\int dxde\, J_z(x,e)\pp_z\ln P^{\rm ss}(x,e)$ denotes the rate of change of the Shannon entropy, while $\bar{Q}_z=\int dxde\, J_z(x,e)F_z(x,e)$ denotes the heat flow from $z$ into the environment.
Note that $F_x(x,e)=f(x,x+e)$ and $F_e(x,e)=-e$ are the forces and $J_z(x,e)=F_z(x,e)P^{\rm ss}(x,e)-D_z\pp_zP^{\rm ss}(x,e)$ is the probability current in the corresponding Fokker--Planck equation.
The flow of information from $e$ to $x$,
\begin{equation}
\bar{I}_{e\to x}=-\int dxde\, J_e(x,e)\pp_e\ln\frac{P^{\rm ss}(x,e)}{P^{\rm ss}(x)P^{\rm ss}(e)},
\end{equation}
is (minus) the variation of the mutual information between $x$ and $e$ \cite{Horowitz.2014.NJP}
\begin{equation}
I(x;e)=\int dxde\,P^{\rm ss}(x,e)\ln\frac{P^{\rm ss}(x,e)}{P^{\rm ss}(x)P^{\rm ss}(e)}.
\end{equation}
Since the noise is in equilibrium, we have that $\int dx\,J_e(x,e)=0$.
Consequently, we obtain that $\bar{Q}_e=0$ and $\bar{I}_{e\to x}=\bar{S}_e$.
Thus, $\bar{\sigma}_e=\bar{I}_{e\to x}\ge 0$ and $\bar{\sigma}=\bar{\sigma}_x+\bar{I}_{e\to x}$.
This implies that, in addition to the system entropy production, $\bar{\sigma}_x$, there is a positive information flow, $\bar{I}_{e\to x}$, from the controller into the system, corresponding to the information contribution in $\avg{\sigma}$.

As regards Langevin systems [Eq.~\eqref{eq:Langevin.feedback}] involving continuous measurements, a tighter bound can be obtained; more specifically, for arbitrary time-integrated currents being $\mca{O}[\bX]=\int_0^{\mca{T}}dt\,\Lambda(x)\circ\dot{x}$, where $\Lambda(x)$ is an arbitrary projection function, we can prove that
\begin{equation}
\frac{\mrm{Var}[\mca{O}]}{\avg{\mca{O}}^2}\ge\frac{2}{\avg{\sigma}}=\frac{2}{\mca{T}(\bar{\sigma}_x+\bar{I}_{e\to x})}.\label{eq:main.TUR.cont}
\end{equation}
This lower bound is analogous to the conventional one [Eq.~\eqref{eq:TUR.conv}] and tighter than that in Eq.~\eqref{eq:main.UR}.
The result implies that the fluctuation is bounded not only by entropy production, $\bar{\sigma}_x$, but also by information flow, $\bar{I}_{e\to x}$.
The larger the information contribution is, the higher the precision of the observables.
When there is no feedback to the system, $\bar{I}_{e\to x}=0$, and $\avg{\sigma}$ becomes the system entropy production.
The detailed derivation of Eq.~\eqref{eq:main.TUR.cont} is provided in \ref{app:deriv.TUR.cont}.

\section{Example}\label{sect:exam}

We apply the derived uncertainty relation to study the precision of a flashing ratchet, which is a model of Brownian ratchet.
First, we describe the conception of the flashing ratchet in the presence of external controller that utilises information obtained from the measurements to rectify a directed motion.
Afterward, we use both discrete- and continuous-state models of flashing ratchet to validate the derived bound.
Both the continuous- and discrete-time ratchets are considered in the discrete-state system.

Let us introduce a flashing ratchet comprising of an overdamped Brownian particle in contact with an equilibrium heat bath at temperature $T$.
The particle evolves under an external asymmetric potential $V(x)$, which can be either on or off depending on the feedback control.
The dynamics of the particle can be described by the Langevin equation:
\begin{equation}
\gamma\dot{x}=\lambda(x)F(x)+\xi,
\end{equation}
where $x$ denotes the position of the particle, $\gamma$ denotes the friction coefficient and $\xi$ denotes white Gaussian noise with zero mean and time correlation $\avg{\xi(t)\xi(t')}=2\gamma T\delta(t-t')$.
The force is given by $F(x)=-\pp_xV(x)$, where $V(x)$ is periodic with $V(x+L)=V(x)$, and $L$ denotes the period of the potential.
The term $\lambda(x)$ denotes a control protocol that takes value $1$ or $0$, corresponding to switching on or off the potential.

In the previous conducted studies \cite{Cao.2004.PRL,Dinis.2005.EPL}, the protocol $\lambda(x)$ can be determined by $\lambda(x)=\Theta(F(x))$, where $\Theta(z)$ denotes the Heaviside function given by $\Theta(z)=1$ if $z>0$ and $0$ otherwise.
This indicates that the potential is turned on only when the net force applied to the particles is positive.
The measurements in these studies were assumed to be perfect, i.e. there is no error in the measurement outcome of the sign of $F(x)$.
This feedback control strategy was shown to be the best possible strategy for maximising the average velocity of one particle.
However, it is not the best strategy for collective flashing ratchet, where more than one particle exists.
Taking a more realistic model into account, Refs. \cite{Feito.2007.EPJB,Cao.2009.PA} studied the flashing ratchet with imperfect measurement.
The error in the estimation of the sign of $F(x)$ occurs with a probability $r\in[0,0.5]$.
Equivalently, the potential is switched wrongly with probability $r$, i.e. the potential can be turned off when $F(x)>0$ or turn on when $F(x)\le0$ with probability $r$.

\subsection{Discrete-measurement and discrete-state case}

In all the studies discussed so far, measurements are continuously executed, which is difficult to implement from the perspective of experimental realisation.
Moreover, there is a redundancy in the information that is obtained from continuous measurements; thus, leading to an inefficient implementation from the perspective of energetic cost.
We consider a discrete-state flashing ratchet with discretely repeated measurements and feedback control in what follows.

\subsubsection{Continuous-time model}
\begin{figure}[t]
\centering
\includegraphics[width=12cm]{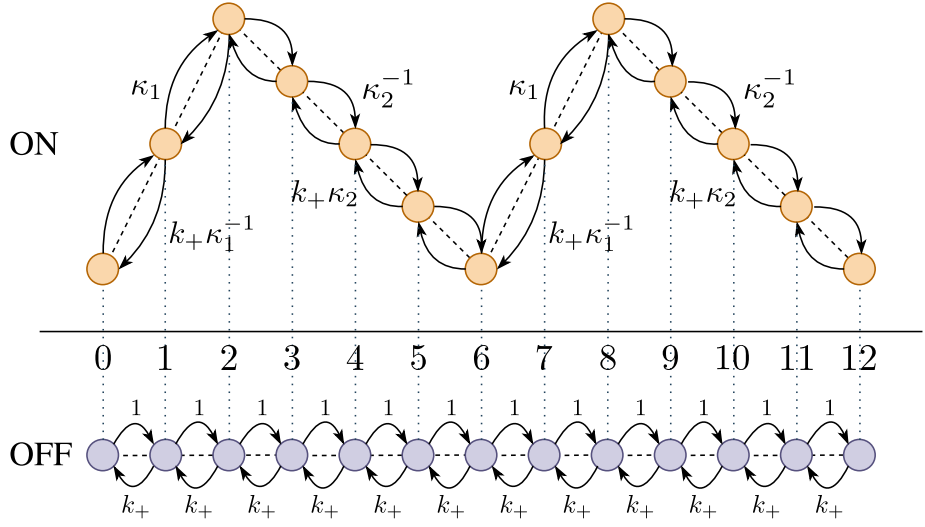}
\protect\caption{Illustration of discrete-state flashing ratchet, where $\mca{N}=6,~\mca{N}_1=2$ and $\mca{N}_2=4$. In the presence of the ratchet potential, the particle transits between adjacent states with predetermined rates. However, when the potential is off, the ratchet obeys a random walk with forward and backward transition rates equal to $1$ and $k_{+}$, respectively.}\label{fig:flashing.ratchet}
\end{figure}

We consider a one-dimensional discrete-state flashing ratchet studied without using feedback control in Refs.~\cite{Freund.1999.PRE,Zhou.2004.PA}.
The ratchet comprises a Brownian particle and has discrete states $n$ located at position $n\Delta x~(n\in\mbb{Z})$, where $\Delta x$ denotes the distance between neighbouring states.
The particle is only allowed to jump between adjacent states, i.e. the particle cannot instantly transit from state $m$ to state $n$, for $|m-n|>1$.
The periodic potential is approximated by $\mca{N}=\mca{N}_1+\mca{N}_2$ states, as illustrated in Fig.~\ref{fig:flashing.ratchet}.
For each $n\in\mbb{Z}$, $\overline{n}\equiv n~(\mrm{mod}~\mca{N})$ is defined as the remainder of the Euclidean division of $n$ by $\mca{N}$.
Suppose that the particle is in state $n$, then the potential should be turned off if $0\le\overline{n}<\mca{N}_1$ and turned on otherwise, i.e. if $\mca{N}_1\le\overline{n}<\mca{N}$.
This is an ideal control protocol, which maximises the instant velocity of the particle.
Moreover, when the measurement is performed, there exists an error due to the noise and the potential is switched wrongly with a probability $r\in(0,1)$.
Particularly, the conditional probability that characterises the measurement error is given as follows:
\begin{equation}
p(s|x)=\begin{cases}
r, & \text{if}~s=1~\text{and}~0\le\overline{x}<\mca{N}_1,\\
r, & \text{if}~s=0~\text{and}~\mca{N}_1\le\overline{x}<\mca{N},\\
1-r, & \text{otherwise},
\end{cases}
\end{equation}
where $s=1$ and $s=0$ show that the potential is switched on and off, respectively, while $x$ denotes the system state when executing the measurement.

When the potential is on, the transition rate $\Gamma_{n,m}$ from state $m$ to state $n$ is given by
\begin{equation}
\begin{aligned}
\Gamma_{n+1,n}&=\kappa_1,~\Gamma_{n,n+1}=k_{+}\kappa_1^{-1},~\forall n:\overline{n}=0,\dots,\mca{N}_1-1,\\
\Gamma_{n+1,n}&=\kappa_2^{-1},~\Gamma_{n,n+1}=k_{+}\kappa_2,~\forall n:\overline{n}=\mca{N}_1,\dots,\mca{N}-1,\\
\Gamma_{n,m}&=0,~\forall~|m-n|>1.
\end{aligned}
\end{equation}
Herein, $k_{+}>0$ reflects the asymmetry in transitions due to a load force, $V_{\rm max}$ denotes the peak of the potential, and
\begin{equation}
\kappa_1=\exp\bras{-\frac{V_{\rm max}}{2\mca{N}_1k_{\rm B}T}},~\kappa_2=\exp\bras{-\frac{V_{\rm max}}{2\mca{N}_2k_{\rm B}T}}.
\end{equation}
When $k_{+}=1$, i.e. there is no load force, the transition rates satisfy the local detailed balance
\begin{equation}
\frac{\Gamma_{n+1,n}}{\Gamma_{n,n+1}}=\exp\bra{\frac{V_n-V_{n+1}}{k_{\rm B}T}},
\end{equation}
where $V_n$ denotes the potential at state $n$, given by
\begin{equation}
V_n=\begin{cases}
V_{\rm max}\overline{n}/\mca{N}_1, & \text{if}~ \overline{n}=0,\dots,\mca{N}_1-1,\\
V_{\rm max}(\mca{N}-\overline{n})/\mca{N}_2, & \text{if}~ \overline{n}=\mca{N}_1,\dots,\mca{N}-1.
\end{cases}
\end{equation}
Hereafter, we set $k_{\rm B}T=1$.
In the continuous limit, i.e. $\mca{N}\to\infty$, the discrete potential converges to the following continuous sawtooth potential:
\begin{equation}\label{eq:saw.potential}
V(x)=\begin{cases}
V_{\rm max}x/(aL), & \text{if}~0\le x\le aL,\\
V_{\rm max}(L-x)/\bras{(1-a)L}, & \text{if}~aL<x\le L,
\end{cases}
\end{equation}
where $a=\lim_{\mca{N}\to\infty}\bra{\mca{N}_1/\mca{N}}$ is a given constant.
Let $P_n(t)$ denote the probability of the system at state $n$ and time $t$.
Thus, the probability distribution is governed by the master equation
\begin{equation}
\pp_tP_n(t)=\sum_{m}\bras{\Gamma_{n,m}P_m(t)-\Gamma_{m,n}P_n(t)}.\label{eq:master.equation}
\end{equation}
When the potential is off, the dynamics of the particle becomes a continuous-time random walk with forward and backward transition rates equal to $1$ and $k_{+}$, respectively.

Let $\bms{P}(x_0,0,s;x_1,\Delta t)$ denote the probability that the system is at state $x_0$ at time $t=0$ with the measurement outcome $s$ and being in state $x_1$ at time $t=\Delta t$.
Since the system is periodic, we define the probability distribution $\bms{Q}(x_0,0,s;x_1,\Delta t)$ for $x_0\in[0,\mca{N}-1]$ and $x_1\in\mbb{Z}$ as follows:
\begin{equation}
\bms{Q}(x_0,0,s;x_1,\Delta t)=\sum_{m,n}\bms{P}(n,0,s;m,\Delta t)\delta_{\overline{n},x_0}\delta_{m-n,x_1-x_0}.
\end{equation}
We note that $\bms{Q}(x_0,0,s;x_1,\Delta t)$ is normalised, i.e. $\sum_{x_0=0}^{\mca{N}-1}\sum_{x_1\in\mbb{Z}}\sum_{s=0}^{1}\bms{Q}(x_0,0,s;x_1,\Delta t)=1$.
The average of the system entropy production is equal to zero since the system is in the steady state.
Therefore, the quantity $\avg{\sigma}$ can be evaluated as
\begin{equation}\label{eq:sigma.avg}
\avg{\sigma}=N\avgl{\ln\frac{w(x_1,\Delta t|x_0,0,s)}{w(x_0,\Delta t|x_1,0,s)}+\ln\frac{p(s|x_0)}{p(s|x_1)}}_{\bms{Q}},
\end{equation}
where the average is taken with respect to the probability distribution $\bms{Q}(x_0,0,s;x_1,\Delta t)$.

\subsubsection{Discrete-time model}

Let us consider a discrete-time model of flashing ratchet, where the control protocol is the same as that of the continuous-time model.
Its dynamics is described by a Markov chain
\begin{equation}
P_n(t+\tau)=\sum_{m}\Lambda_{n,m}P_m(t),
\end{equation}
where $\tau$ denotes the time step and $\Lambda_{n,m}$ denotes the transition probability from state $m$ to $n$.
To be consistent with the continuous-time model, the probability $\Lambda_{n,m}$ is set as follows.
The ratchet transits between states with the following probabilities when the potential is on are given by
\begin{equation}
\Lambda_{n,m}=\begin{cases}
\tau\Gamma_{n,m}, & \text{if}~m\neq n,\\
1-\tau\bra{\Gamma_{n+1,n}+\Gamma_{n-1,n}}, & \text{if}~m=n.
\end{cases}
\end{equation}
When the potential is off, the ratchet becomes a discrete-time random walk with the transition probabilities given by $\Lambda_{n,n-1}=\tau,~\Lambda_{n,n+1}=\tau k_{+},~\Lambda_{n,n}=1-\tau(1+k_{+})$ and $\Lambda_{n,m}=0$ for all $|m-n|>1$.
Moreover, the time step must be properly chosen to ensure the positivity of the transition probabilities, i.e.
\begin{equation}
\tau\le\min\brab{\frac{1}{1+k_{+}},\frac{1}{\max_{n}\bras{\Gamma_{n+1,n}+\Gamma_{n-1,n}}}}.
\end{equation}
In addition, the gap time between two consecutive measurements should be a multiple of time step, i.e. $\Delta t/\tau\in\mbb{N}$.
The term $\avg{\sigma}$ can be evaluated analogously as in Eq.~\eqref{eq:sigma.avg}.
\begin{figure}[t]
\centering
\includegraphics[width=10cm]{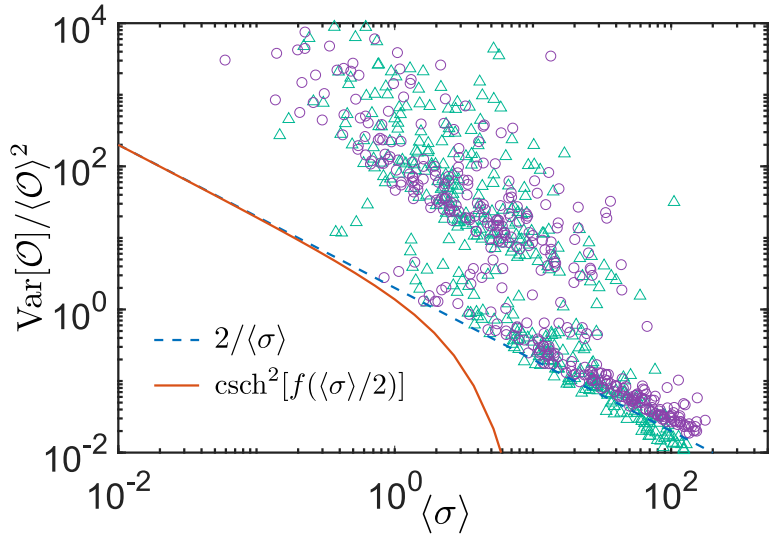}
\protect\caption{Numerical verification of the derived uncertainty relation in the discrete-state flashing ratchet system. The circular and triangular points denote the simulation results of the continuous- and discrete-time models, respectively. $\csch^2[f(\avg{\sigma}/2)]$ and $2/\avg{\sigma}$ are depicted by solid and dashed lines, respectively. The parameter ranges are $\mca{N}\in[3,30],~\mca{N}_1\in[1,\mca{N}/2],~V_{\rm max}\in[0.1,10],~k_{+}\in[0.1,10]$ and $r\in(0,0.5)$. The remaining parameters are $\Delta t\in[0.01,1],~\mca{T}\in[2,10]$ in continuous-time model, while $\tau\in[0.1,0.5],~\Delta t/\tau\in[1,10],~\mca{T}/\tau\in[10,100]$ in discrete-time model.}\label{fig:example.results.1}
\end{figure}
\subsubsection{Bound on the precision of the discrete-state ratchet}
Now, we verify the derived bound for the following observable:
\begin{equation}
\bO[\bX]=x_N-x_0.
\end{equation}
This observable is a current, which represents the distance travelled by the particle.
The relative fluctuation, $\mrm{Var}[\bO]/\avg{\bO}^2$, reflects the precision of the ratchet.
According to Eq.~\eqref{eq:main.UR}, the inequality $\mrm{Var}[\bO]/\avg{\bO}^2\ge\csch^2[f(\avg{\sigma}/2)]$ should be satisfied.
We conduct stochastic simulations for both the continuous- and discrete-time models and numerically evaluate the precision, $\mrm{Var}[\bO]/\avg{\bO}^2$ and the bound term, $\avg{\sigma}$.
For each random parameter setting, $(\mca{N}_1,\mca{N},V_{\rm max},k_{+},r,\Delta t,\mca{T})$, we collect $10^7$ realisations for the calculation.
The ranges of the parameters are shown in the caption of Fig.~\ref{fig:example.results.1}.
We plot $\mrm{Var}[\bO]/\avg{\bO}^2$ as a function of $\avg{\sigma}$ in Fig.~\ref{fig:example.results.1}, where the circular and triangular points represent the results of the continuous- and discrete-time models, respectively.
We depict the saturated case of the derived bound and the conventional bound by solid and dashed lines, respectively.
As shown in Fig.~\ref{fig:example.results.1}, all the points are located above the solid line; thus, the validity of the derived bound is empirically verified.
However, several triangular points lie below the dashed line, which implies that the conventional bound is violated.
\begin{figure}[t]
\centering
\includegraphics[width=15cm]{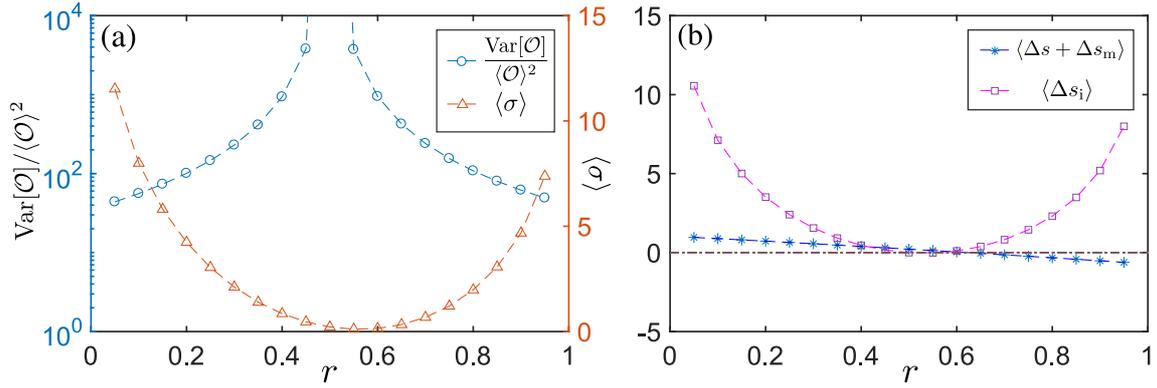}
\protect\caption{(a) The uncertainty and the total entropy production $\avg{\sigma}$ corresponding to the measurement error in the discrete-time model. Uncertainty and entropy production are depicted by circular and triangular points, respectively. (b) The entropy productions, $\avg{\Delta s+\Delta s_{\mrm{m}}}$ and $\avg{\Delta s_{\mrm{i}}}$, as functions of the measurement error. $\avg{\Delta s+\Delta s_{\mrm{m}}}$ and $\avg{\Delta s_{\mrm{i}}}$ are depicted by star and square points, respectively. The measurement error, $r$, is varied from $0.05$ to $0.95$, while the remaining parameters are fixed: $\mca{N}=4,~\mca{N}_1=2,V_{\rm max}=1,~k_{+}=1,~\Delta t=0.35,~\mca{T}=3.5$ and $\tau=0.35$.}\label{fig:example.results.2}
\end{figure}

We plot the uncertainty in the observable and the total entropy production as functions of measurement error parameter, $r$, in Fig.~\ref{fig:example.results.2}(a).
When $r$ decreases to $0$ or increases to $1$, more information is obtained from measurement.
Therefore, this results in higher entropy production and lower uncertainty.
It is interesting that the uncertainty in observable declines exponentially when $r$ is either decreased from $0.5$ to $0$ or increased from $0.5$ to $1$.
When $r$ is increased to $1$, the error in the measurements leads to a reverse motion to the left side.
In Fig.~\ref{fig:example.results.2}(b), we plot $\avg{\Delta s+\Delta s_{\rm m}}$ and $\avg{\Delta s_{\rm i}}$ as functions of $r$.
As seen, $\avg{\Delta s+\Delta s_{\rm m}}$ becomes negative when $r\ge 0.65$, while $\avg{\sigma}$ is always positive.
This implies that in this case, the controller works as a kind of Maxwell's demon.

\subsection{Continuous-measurement and continuous-state case}

\begin{figure}[t]
\centering
\includegraphics[width=10cm]{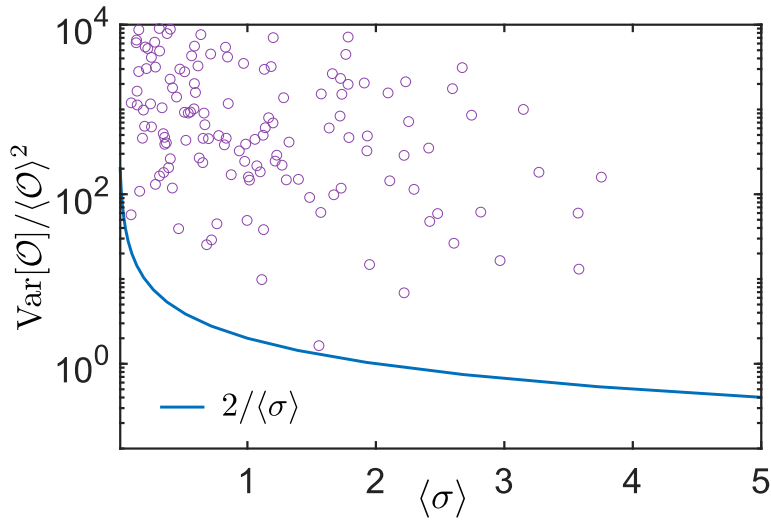}
\protect\caption{Numerical verification of the uncertainty relation in the continuous-state flashing ratchet. $\mrm{Var}[\bO]/\avg{\bO}^2$ and $2/\avg{\sigma}$ are represented by the circles and the solid line, respectively. The parameter ranges are $V_{\rm max}\in[0.5,2],~a\in[0.1,0.5],~D_x,D_e\in[10^{-3},10^0]$ and $\mca{T}\in[0.1,2]$. The remaining parameters are fixed: $\gamma=1$ and $L=1$.}\label{fig:example.results.3}
\end{figure}

Finally, we consider a continuous-state flashing ratchet under continuous measurement.
Its dynamics is governed by the following equations:
\begin{equation}
\begin{aligned}
\gamma\dot{x}&=\lambda(m)F(x)+\xi,\\
m&=x+e.
\end{aligned}
\end{equation}
Here, $F(x)=-\pp_{x}V(x)$ [the form of $V(x)$ is given in Eq.~\eqref{eq:saw.potential}], $e$ is the OU process defined as in Eq.~\eqref{eq:OU.proc} and the protocol $\lambda(m)$ is a periodic function, i.e. $\lambda(m)=\lambda(m+L)$, defined as
\begin{equation}
\lambda(m)=\begin{cases}
0, & \text{if}~0\le m\le aL,\\
1, & \text{if}~aL<m<L.
\end{cases}
\end{equation}
We verify the bound derived in Eq.~\eqref{eq:main.TUR.cont} with the current $\mca{O}[\bX]=\int_0^{\mca{T}}dt\,\dot{x}=x(\mca{T})-x(0)$, which expresses the ratchet displacement.
According to Eq.~\eqref{eq:main.TUR.cont}, the inequality $\mrm{Var}[\bO]/\avg{\bO}^2\ge 2/\avg{\sigma}$ should be satisfied.
We randomly sample the parameters and run computer simulations to evaluate $\avg{\bO}$, $\mrm{Var}[\bO]$ and $\avg{\sigma}$.
For each parameter setting, we collect $10^7$ trajectories by using the Euler--Maruyama method with the time step $\Delta t=10^{-4}$.
$\mrm{Var}[\bO]/\avg{\bO}^2$ is plotted as a function of $\avg{\sigma}$ in Fig.~\ref{fig:example.results.3}.
Since all the points are located above the line, the derived bound is empirically verified.

The ratchet precision is evaluated also by the Peclet number \cite{Freund.1999.PRE}, which is defined as follows:
\begin{equation}
\Pe=\frac{\avg{v}}{\bms{D}},\label{eq:Peclet.num.def}
\end{equation}
where $\avg{v}$ and $\bms{D}$ are, respectively, the mean velocity and the effective diffusion coefficient of the ratchet.
Substituting
\begin{equation}
\avg{v}=\lim_{\mca{T}\to\infty}\frac{\avg{\bO}}{\mca{T}},~ \bms{D}=\lim_{\mca{T}\to\infty}\frac{\mrm{Var}[\bO]}{2\mca{T}}
\end{equation}
into Eq.~\eqref{eq:Peclet.num.def}, we obtain
\begin{equation}\label{eq:Pe.Fano.inv}
\Pe=\lim_{\mca{T}\to\infty}\frac{2\avg{\bO}}{\mrm{Var}[\bO]},
\end{equation}
which indicates that $\Pe$ is proportional to the inverse of the Fano factor: the larger $\Pe$, the higher the ratchet precision.
From the derived bound, we can readily obtain the following upper bound on $\Pe$:
\begin{equation}\label{eq:peclet.ine}
\Pe\le\frac{\avg{\sigma}}{\avg{\bO}}.
\end{equation}
This inequality can be rewritten as $\Pe\times\avg{\bO}\le\avg{\sigma}$, which implies a trade-off between the ratchet precision and the distance travelled, i.e. with a fixed energy cost $\avg{\sigma}$, a ratchet cannot attain both a high precision and a long displacement.

\section{Conclusion}
We derived the uncertainty relation for steady-state systems involving repeated measurements and feedback control.
We showed that the relative fluctuation of arbitrary observables that are antisymmetric under time reversal is constrained from below by $\avg{\sigma}$, which is the sum of the entropy production and the mutual information.
For Langevin dynamics involving continuous measurement, we also demonstrated a tighter bound from below on the fluctuation of time-integrated currents.
Then, we empirically validated the derived bound for the displacement of a flashing ratchet.

The bound for the discrete-measurement case was derived from the fluctuation theorem, which holds for both continuous- and discrete-time systems.
Although the measurements were performed discretely, we did not observe any violation of the conventional bound in the stochastic simulations for the continuous-time ratchet, i.e. $\mrm{Var}[\bO]/\avg{\bO}^2\ge 2/\avg{\sigma}$ held for all the parameter settings in the continuous-time model.
We remark that proving this inequality could significantly improve the bound and, thus, requires further investigation.\\\\
\noindent\emph{Note added}:
We note that an independent related result has been obtained in Ref.~\cite{Potts.2019.PRE}, where the authors derive an uncertainty relation for systems involving measurement and feedback control. The relation holds even when the time-symmetry is broken (i.e. $\mca{P}_{\rm F}\neq\mca{P}_{\rm R}$) and includes not only thermodynamic quantities in the forward experiment but also those in the backward experiment. When $\mca{P}_{\rm F}=\mca{P}_{\rm R}$, the lower bound reduces to $2/(e^{\avg{\sigma}}-1)$, which is not tight as Eqs.~\eqref{eq:main.UR} and \eqref{eq:main.TUR.cont}.

\section*{Acknowledgement}
This work was supported by Ministry of Education, Culture, Sports, Science and Technology (MEXT) KAKENHI Grants No. JP16K00325 and No. JP19K12153.

\appendix

\section{Derivation of the uncertainty relation for systems including discrete measurement and feedback control}\label{app:deriv.TUR.disc}
First, we note that the joint probability distribution of $\sigma$ and $\bO$, $P(\sigma,\bO)$, obeys the strong DFT
\begin{equation}\label{eq:strong.DFT.sig.obs}
P(\sigma,\bO)=e^{\sigma}P(-\sigma,-\bO).
\end{equation}
Equation \eqref{eq:strong.DFT.sig.obs} can be readily obtained as follows:
\begin{equation}
\begin{aligned}
&P(\sigma,\bO)\\
&=\int\mca{D}\bZ\,\delta(\sigma-\sigma(\bX,\bM))\delta(\bO-\bO[\bX])\mca{P}(\bX,\bM)\\
&=\int\mca{D}\bZ\,\delta(\sigma-\sigma(\bX,\bM))\delta(\bO-\bO[\bX])e^{\sigma(\bX,\bM)}\mca{P}(\bXr,\bMr)\\
&=e^{\sigma}\int\mca{D}\bZ\,\delta(\sigma-\sigma(\bX,\bM))\delta(\bO-\bO[\bX])\mca{P}(\bXr,\bMr)\\
&=e^{\sigma}\int\mca{D}\bZr\,\delta(\sigma+\sigma(\bXr,\bMr))\delta(\bO+\bO[\bXr])\mca{P}(\bXr,\bMr)\\
&=e^{\sigma}P(-\sigma,-\bO).
\end{aligned}
\end{equation}
Here, $\mca{D}\bZ\equiv\mca{D}\bX\mca{D}\bM$ and $\mca{D}\bZr\equiv\mca{D}\bXr\mca{D}\bMr$.
Inspired by Ref.~\cite{Merhav.2010.JSM}, where the statistical properties of entropy production were obtained from the strong DFT, we derive the uncertainty relation mainly from Eq.~\eqref{eq:strong.DFT.sig.obs}.
By observing that
\begin{align}
1&=\intf d\sigma\intf d\bO\, P(\sigma,\bO)\nonumber\\
&=\inth d\sigma\intf d\bO\,(1+e^{-\sigma})P(\sigma,\bO),
\end{align}
we introduce a probability distribution $Q(\sigma,\bO)\equiv(1+e^{-\sigma})P(\sigma,\bO)$, which is defined over $[0,\infty)\times(-\infty,\infty)$.
The first and second moments of $\sigma$ and $\bO$ can be expressed with respect to the distribution $Q(\sigma,\bO)$ as follows:
\begin{equation}
\begin{aligned}
\avg{\sigma}&=\avgl{\sigma\tanh\bra{\frac{\sigma}{2}}}_{Q},~\avg{\sigma^{2}}=\avgl{\sigma^{2}}_{Q},\\
\avg{\bO}&=\avgl{\bO\tanh\bra{\frac{\sigma}{2}}}_{Q},~\avg{\bO^{2}}=\avgl{\bO^{2}}_{Q},
\end{aligned}
\end{equation}
where $\avg{\dots}_{Q}$ denotes the expectation with respect to $Q(\sigma,\bO)$.
By applying the Cauchy--Schwartz inequality to $\avg{\bO}$, we obtain
\begin{equation}
\avg{\bO}^2=\avgl{\bO\tanh\bra{\frac{\sigma}{2}}}^2_{Q}\le\avg{\bO^2}_{Q}\avgl{\tanh^2\bra{\frac{\sigma}{2}}}_{Q}.\label{eq:bound1}
\end{equation}
The last term in the right-hand side of Eq.~\eqref{eq:bound1} can be further upper bounded.
We observe that
\begin{equation}\label{eq:bound2}
\begin{aligned}
\avgl{\tanh^2\bra{\frac{\sigma}{2}}}_{Q}&=\avgl{\tanh^2\bras{f\bra{\frac{\sigma}{2}\tanh\bra{\frac{\sigma}{2}}}}}_{Q}\\
&\le\tanh^2\bras{f\bra{\avg{\sigma}/2}}.
\end{aligned}
\end{equation}
The equality in Eq.~\eqref{eq:bound2} is obtained from the fact that $f(x)$ is the inverse function of $x\tanh(x)$.
The inequality in Eq.~\eqref{eq:bound2} can be obtained as follows.
First, we show that $\chi(x)=\tanh^2[f(x)]$ is a concave function over $[0,+\infty)$.
Indeed, using the relation $f(x)\tanh\bras{f(x)}=x$ and performing simple calculations, we obtain
\begin{equation}
\frac{d^2\chi(x)}{dx^2}=\frac{4\bra{4f(x)-\sinh\bras{4f(x)}}}{\bra{2f(x)+\sinh\bras{2f(x)}}^3}.
\end{equation}
Since $4f(x)\le\sinh\bras{4f(x)}$, we have $d^2\chi(x)/dx^2\le 0,~\forall\,x\ge 0$; thus, implying that $\tanh^2[f(x)]$ is a concave function.
Applying Jensen's inequality to this function, we obtain
\begin{equation}
\begin{aligned}
&\avgl{\tanh^2\bras{f\bra{\frac{\sigma}{2}\tanh\bra{\frac{\sigma}{2}}}}}_{Q}\\
&\le\tanh^2\bras{f\bra{\avgl{\frac{\sigma}{2}\tanh\bra{\frac{\sigma}{2}}}_{Q}}}=\tanh^2\bras{f\bra{\avg{\sigma}/2}}.
\end{aligned}
\end{equation}
From Eqs.~\eqref{eq:bound1} and \eqref{eq:bound2}, we have
\begin{equation}
\avg{\bO}^2\le\avg{\bO^2}\tanh^2\bras{f\bra{\avg{\sigma}/2}}.\label{eq:temp.bound}
\end{equation}
By transforming Eq.~\eqref{eq:temp.bound}, we obtain the derived bound [Eq.~\eqref{eq:main.UR}] for the observable $\bO$.

\section{Derivation of the uncertainty relation for systems involving continuous measurement and feedback control}\label{app:deriv.TUR.cont}
Let $P(x,e,t)$ be the probability distribution function of the joint system [Eqs.~\eqref{eq:Langevin.feedback} and \eqref{eq:OU.proc}].
Its time evolution is described by the Fokker--Planck equation as follows:
\begin{equation}
\pp_tP(x,e,t)=-\pp_xJ_x(x,e,t)-\pp_eJ_e(x,e,t),
\end{equation}
where $J_x(x,e,t)=f(x,x+e)P(x,e,t)-D_{x}\pp_xP(x,e,t)$ and $J_e(x,e,t)=-eP(x,e,t)-D_e\pp_eP(x,e,t)$ are probability currents.
Hereafter, we focus exclusively on the nonequilibrium steady state, for which the probability distribution and currents are $P^{\rm ss}(x,e)$ and $\bm{J}^{\rm ss}(x,e)\equiv[J_x^{\rm ss}(x,e),J_e^{\rm ss}(x,e)]^\top$, respectively.

We use the Cram{\'e}r--Rao inequality \cite{Hasegawa.2019.PRE} and the perturbation technique \cite{Dechant.2018.arxiv} to derive the uncertainty relation for the systems under consideration.
Let us consider an auxiliary dynamics described by
\begin{equation}
\begin{aligned}
\dot{x}&=f(x,x+e)+\theta\frac{J_x^{\rm ss}(x,e)}{P^{\rm ss}(x,e)}+\xi,\\
\dot{e}&=-e+\theta\frac{J_e^{\rm ss}(x,e)}{P^{\rm ss}(x,e)}+\eta,
\end{aligned}
\end{equation}
where $\theta$ is a perturbation parameter.
When $\theta=0$, this auxiliary dynamics becomes the original one.
Let $P_\theta^{\rm ss}(x,e)$ be the stationary distribution of this auxiliary dynamics; $P_\theta^{\rm ss}(x,e)=P^{\rm ss}(x,e)$ can be easily confirmed.
The probability current of the auxiliary dynamics is scaled as follows:
\begin{equation}
J_{\theta,x}^{\rm ss}(x,e)=(1+\theta)J_{x}^{\rm ss}(x,e).
\end{equation}
Let $\bX=\set{x(t)}_{t=0}^{t=\mca{T}}$ and $\bE=\set{e(t)}_{t=0}^{t=\mca{T}}$ be the trajectories of the system states and the noise, respectively.
In the auxiliary dynamics, the path probability using the Ito discretisation is expressed as
\begin{equation}
\begin{aligned}
\mca{P}_\theta(\bX,\bE)&\propto P^{\rm ss}_\theta(x(0),e(0))\exp\bra{-\frac{1}{4D_x}\int_0^{\mca{T}}dt\bras{\dot{x}-f(x,x+e)-\theta\frac{J_x^{\rm ss}(x,e)}{P^{\rm ss}(x,e)}}^2}\\
&\times\exp\bra{-\frac{1}{4D_e}\int_0^{\mca{T}}dt\bras{\dot{e}+e-\theta\frac{J_e^{\rm ss}(x,e)}{P^{\rm ss}(x,e)}}^2}.
\end{aligned}
\end{equation}
For an arbitrary function $\phi(\bX)$, we define $\avg{\phi}_\theta=\int\mca{D}\bX\mca{D}\bE\,\mca{P}_\theta(\bX,\bE)\phi(\bX)$ and $\mrm{Var}_\theta[\phi]=\avg{(\phi-\avg{\phi}_\theta)^2}_\theta$.
Since $\avg{\mca{O}}_\theta$ is a function of $\theta$, we can consider $\mca{O}$ as one of its estimators.
According to the Cram{\'e}r--Rao inequality, the precision of this estimator is lower bounded by the Fisher information $\mca{I}(\theta)$ as follows:
\begin{equation}\label{eq:Cramer.Rao.ine}
\frac{\mrm{Var}_\theta[\mca{O}]}{\bra{\pp_\theta\avg{\mca{O}}_\theta}^2}\ge\frac{1}{\mca{I}(\theta)},
\end{equation}
where $\mca{I}(\theta)\equiv-\avg{\pp_\theta^2\ln\mca{P}_\theta(\bX,\bE)}_\theta$.
Since $\avg{\mca{O}}_\theta=\mca{T}\int dx\int de\,\Lambda(x)J_{\theta,x}^{\rm ss}(x,e)$, we have $\avg{\mca{O}}_\theta=(1+\theta)\avg{\mca{O}}$ and, thus, $\pp_\theta\avg{\mca{O}}_\theta=\avg{\mca{O}}$.
Moreover, through some algebraic calculations, we obtain
\begin{equation}
\mca{I}(0)=\frac{\mca{T}}{2}\int dx\int de\,\frac{\bm{J}^{\rm ss}(x,e)^\top\bm{D}^{-1}\bm{J}^{\rm ss}(x,e)}{P^{\rm ss}(x,e)}=\frac{1}{2D_x}\avgl{\int_0^{\mca{T}}dt\,f(x,x+e)\circ\dot{x}},
\end{equation}
where $\bm{D}\equiv\mrm{diag}(D_x,D_e)\in\mbb{R}^{2\times2}$.
As shown, the Fisher information is directly proportional to $\avg{\sigma}$, i.e. $\mca{I}(0)=\avg{\sigma}/2$.
By letting $\theta=0$ in Eq.~\eqref{eq:Cramer.Rao.ine}, we readily obtain
\begin{equation}
\frac{\mrm{Var}[\mca{O}]}{\avg{\mca{O}}^2}\ge\frac{2}{\avg{\sigma}}.
\end{equation}

\section*{References}
%\bibliographystyle{iopart-num.bst}
%\bibliography{refs}

\begin{thebibliography}{66}
\expandafter\ifx\csname url\endcsname\relax
  \def\url#1{{\tt #1}}\fi
\expandafter\ifx\csname urlprefix\endcsname\relax\def\urlprefix{URL }\fi
\providecommand{\eprint}[2][]{\url{#2}}

\bibitem{Seifert.2012.RPP}
Seifert U 2012 {\em Rep. Prog. Phys.\/} {\bf 75} 126001

\bibitem{Broeck.2015.PA}
Van~den Broeck C and Esposito M 2015 {\em Physica A\/} {\bf 418} 6--16

\bibitem{Gallavotti.1995.PRL}
Gallavotti G and Cohen E~G~D 1995 {\em Phys. Rev. Lett.\/} {\bf 74}(14)
  2694--2697

\bibitem{Jarzynski.1997.PRL}
Jarzynski C 1997 {\em Phys. Rev. Lett.\/} {\bf 78}(14) 2690--2693

\bibitem{Barato.2015.PRL}
Barato A~C and Seifert U 2015 {\em Phys. Rev. Lett.\/} {\bf 114}(15) 158101

\bibitem{Gingrich.2016.PRL}
Gingrich T~R, Horowitz J~M, Perunov N and England J~L 2016 {\em Phys. Rev.
  Lett.\/} {\bf 116}(12) 120601

\bibitem{Pietzonka.2016.PRE}
Pietzonka P, Barato A~C and Seifert U 2016 {\em Phys. Rev. E\/} {\bf 93}(5)
  052145

\bibitem{Polettini.2016.PRE}
Polettini M, Lazarescu A and Esposito M 2016 {\em Phys. Rev. E\/} {\bf 94}(5)
  052104

\bibitem{Horowitz.2017.PRE}
Horowitz J~M and Gingrich T~R 2017 {\em Phys. Rev. E\/} {\bf 96}(2) 020103

\bibitem{Dechant.2018.JSM}
Dechant A and Sasa S~i 2018 {\em J. Stat. Mech.: Theory Exp.\/} {\bf 2018}
  063209

\bibitem{Proesmans.2017.EPL}
Proesmans K and den Broeck C~V 2017 {\em EPL\/} {\bf 119} 20001

\bibitem{Chiuchiu.2018.PRE}
Chiuchi\`u D and Pigolotti S 2018 {\em Phys. Rev. E\/} {\bf 97}(3) 032109

\bibitem{Barato.2018.NJP}
Barato A~C, Chetrite R, Faggionato A and Gabrielli D 2018 {\em New J. Phys.\/}
  {\bf 20} 103023

\bibitem{Koyuk.2019.JPA}
Koyuk T, Seifert U and Pietzonka P 2019 {\em J. Phys. A: Math. Theor.\/} {\bf
  52} 02LT02

\bibitem{Dechant.2018.JPA}
Dechant A 2018 {\em J. Phys. A: Math. Theor.\/} {\bf 52} 035001

\bibitem{Vu.2019.PRE.Underdamp}
Van~Vu T and Hasegawa Y 2019 {\em Phys. Rev. E\/} {\bf 100}(3) 032130

\bibitem{Vu.2019.PRE.Delay}
Van~Vu T and Hasegawa Y 2019 {\em Phys. Rev. E\/} {\bf 100}(1) 012134

\bibitem{Hasegawa.2019.PRE}
Hasegawa Y and Van~Vu T 2019 {\em Phys. Rev. E\/} {\bf 99}(6) 062126

\bibitem{Proesmans.2019.JSM}
Proesmans K and Horowitz J~M 2019 {\em J. Stat. Mech.: Theory Exp.\/} {\bf
  2019} 054005

\bibitem{Barato.2015.JPCB}
Barato A~C and Seifert U 2015 {\em J. Phys. Chem. B\/} {\bf 119} 6555--6561

\bibitem{Patrick.2016.JSM}
Pietzonka P, Barato A~C and Seifert U 2016 {\em J. Stat. Mech.: Theory Exp.\/}
  {\bf 2016} 124004

\bibitem{Hwang.2018.JPCL}
Hwang W and Hyeon C 2018 {\em J. Phys. Chem. Lett.\/} {\bf 9} 513--520

\bibitem{Pietzonka.2018.PRL}
Pietzonka P and Seifert U 2018 {\em Phys. Rev. Lett.\/} {\bf 120}(19) 190602

\bibitem{Holubec.2018.PRL}
Holubec V and Ryabov A 2018 {\em Phys. Rev. Lett.\/} {\bf 121}(12) 120601

\bibitem{Falasco.2016.PRE}
Falasco G, Pfaller R, Bregulla A~P, Cichos F and Kroy K 2016 {\em Phys. Rev.
  E\/} {\bf 94}(3) 030602

\bibitem{Rotskoff.2017.PRE}
Rotskoff G~M 2017 {\em Phys. Rev. E\/} {\bf 95}(3) 030101

\bibitem{Garrahan.2017.PRE}
Garrahan J~P 2017 {\em Phys. Rev. E\/} {\bf 95}(3) 032134

\bibitem{Gingrich.2017.PRL}
Gingrich T~R and Horowitz J~M 2017 {\em Phys. Rev. Lett.\/} {\bf 119}(17)
  170601

\bibitem{Hyeon.2017.PRE}
Hyeon C and Hwang W 2017 {\em Phys. Rev. E\/} {\bf 96}(1) 012156

\bibitem{Brandner.2018.PRL}
Brandner K, Hanazato T and Saito K 2018 {\em Phys. Rev. Lett.\/} {\bf 120}(9)
  090601

\bibitem{Manikandan.2018.JPA}
Manikandan S~K and Krishnamurthy S 2018 {\em J. Phys. A: Math. Theor.\/} {\bf
  51} 11LT01

\bibitem{Terlizzi.2018.JPA}
Terlizzi I~D and Baiesi M 2018 {\em J. Phys. A: Math. Theor.\/} {\bf 52} 02LT03

\bibitem{Carollo.2019.PRL}
Carollo F, Jack R~L and Garrahan J~P 2019 {\em Phys. Rev. Lett.\/} {\bf
  122}(13) 130605

\bibitem{Shreshtha.2019.EPL}
Shreshtha M and Harris R~J 2019 {\em {EPL}\/} {\bf 126} 40007

\bibitem{Hasegawa.2019.PRL}
Hasegawa Y and Van~Vu T 2019 {\em Phys. Rev. Lett.\/} {\bf 123}(11) 110602

\bibitem{Touchette.2000.PRL}
Touchette H and Lloyd S 2000 {\em Phys. Rev. Lett.\/} {\bf 84}(6) 1156--1159

\bibitem{Touchette.2004.PA}
Touchette H and Lloyd S 2004 {\em Physica A\/} {\bf 331} 140--172

\bibitem{Sagawa.2008.PRL}
Sagawa T and Ueda M 2008 {\em Phys. Rev. Lett.\/} {\bf 100}(8) 080403

\bibitem{Sagawa.2010.PRE}
Sagawa T and Ueda M 2010 {\em Phys. Rev. Lett.\/} {\bf 104}(9) 090602

\bibitem{Horowitz.2011.EPL}
Horowitz J~M and Parrondo J~M~R 2011 {\em {EPL}\/} {\bf 95} 10005

\bibitem{Sagawa.2012.PRE}
Sagawa T and Ueda M 2012 {\em Phys. Rev. E\/} {\bf 85}(2) 021104

\bibitem{Hartich.2014.JSM}
Hartich D, Barato A~C and Seifert U 2014 {\em J. Stat. Mech.: Theory Exp.\/}
  {\bf 2014} P02016

\bibitem{Barato.2014.PRL}
Barato A~C and Seifert U 2014 {\em Phys. Rev. Lett.\/} {\bf 112}(9) 090601

\bibitem{Parrondo.2015.NP}
Parrondo J~M, Horowitz J~M and Sagawa T 2015 {\em Nature Phys.\/} {\bf 11} 131

\bibitem{Potts.2018.PRL}
Potts P~P and Samuelsson P 2018 {\em Phys. Rev. Lett.\/} {\bf 121}(21) 210603

\bibitem{Serreli.2007.N}
Serreli V, Lee C~F, Kay E~R and Leigh D~A 2007 {\em Nature\/} {\bf 445} 523

\bibitem{Lau.2017.MH}
Lau B, Kedem O, Schwabacher J, Kwasnieski D and Weiss E~A 2017 {\em Mater.
  Horiz.\/} {\bf 4}(3) 310--318

\bibitem{Schimansky.1997.PRL}
Schimansky-Geier L, Kschischo M and Fricke T 1997 {\em Phys. Rev. Lett.\/} {\bf
  79}(18) 3335--3338

\bibitem{Freund.1999.PRE}
Freund J~A and Schimansky-Geier L 1999 {\em Phys. Rev. E\/} {\bf 60}(2)
  1304--1309

\bibitem{Cao.2004.PRL}
Cao F~J, Dinis L and Parrondo J~M~R 2004 {\em Phys. Rev. Lett.\/} {\bf 93}(4)
  040603

\bibitem{Craig.2008.AP}
Craig E, Kuwada N, Lopez B and Linke H 2008 {\em Ann. Phys.\/} {\bf 17}
  115--129

\bibitem{Mogilner.1999.EBJ}
Mogilner A and Oster G 1999 {\em Eur. Biophys. J.\/} {\bf 28} 235--242

\bibitem{Siwy.2002.PRL}
Siwy Z and Fuli\'{n}ski A 2002 {\em Phys. Rev. Lett.\/} {\bf 89}(19) 198103

\bibitem{Kundu.2012.PRE}
Kundu A 2012 {\em Phys. Rev. E\/} {\bf 86}(2) 021107

\bibitem{Timpanaro.2019.PRL}
Timpanaro A~M, Guarnieri G, Goold J and Landi G~T 2019 {\em Phys. Rev. Lett.\/}
  {\bf 123}(9) 090604

\bibitem{Shiraishi.2017.arxiv}
Shiraishi N 2017 {\em arXiv preprint\/}  arXiv:1706.00892

\bibitem{Horowitz.2015.JSM}
Horowitz J~M 2015 {\em J. Stat. Mech.: Theory Exp.\/} {\bf 2015} P03006

\bibitem{Horowitz.2014.PRX}
Horowitz J~M and Esposito M 2014 {\em Phys. Rev. X\/} {\bf 4}(3) 031015

\bibitem{Horowitz.2014.NJP}
Horowitz J~M and Sandberg H 2014 {\em New J. Phys.\/} {\bf 16} 125007

\bibitem{Dinis.2005.EPL}
Dinis L, Parrondo J~M~R and Cao F~J 2005 {\em EPL\/} {\bf 71} 536--541

\bibitem{Feito.2007.EPJB}
Feito M and Cao F~J 2007 {\em Eur. Phys. J. B\/} {\bf 59} 63--68

\bibitem{Cao.2009.PA}
Cao F, Feito M and Touchette H 2009 {\em Physica A\/} {\bf 388} 113--119

\bibitem{Zhou.2004.PA}
Zhou Y and Bao J~D 2004 {\em Physica A\/} {\bf 343} 515--524

\bibitem{Potts.2019.PRE}
Potts P~P and Samuelsson P 2019 {\em Phys. Rev. E\/} {\bf 100}(5) 052137

\bibitem{Merhav.2010.JSM}
Merhav N and Kafri Y 2010 {\em J. Stat. Mech.: Theory Exp.\/} {\bf 2010} P12022

\bibitem{Dechant.2018.arxiv}
Dechant A and Sasa S~i 2018 {\em arXiv preprint\/}  arXiv:1804.08250

\end{thebibliography}
\providecommand{\newblock}{}

\end{document}